\documentclass[aps,preprint,amsmath,amssymb]{revtex4}

\usepackage{graphicx}
\usepackage{amsmath}
\usepackage{bm}
\newcommand{\be}{\begin{equation}}
\newcommand{\ee}{\end{equation}}
\newcommand{\bear}{\begin{eqnarray}}
\newcommand{\eear}{\end{eqnarray}}

\begin{document}

\title{Chiral radiative corrections and  $D_s(2317)/D(2308)$ mass 
puzzle}

\author{Taekoon Lee}
\email[]{tlee@kunsan.ac.kr}
\affiliation{Department of Physics, Kunsan National University,
Kunsan 573-701, Korea}
\author{Ian Woo Lee}
\author{D.P. Min}
\affiliation{Department of Physics, Seoul National University,
Seoul 151-742, Korea}

\author{Byung-Yoon Park}
\affiliation{CSSM, University of Adelaide, Adelaide 5005, Australia 
and \\
Department of Physics,Chungnam National University, 
Daejon 305-764, Korea}
%\date{\today}

\begin{abstract}
We show that the one loop chiral corrections for heavy-light mesons
in potential model can explain the small mass
of $D_s(2317)$ as well as
the small mass gap between $D_s(2317)$ and $D(2308)$. 
\end{abstract}

%\pacs{} 
%\keywords{}
%\tighten

\maketitle

The recently observed $D_s(2317)$ \cite{barbar,cleo,belle}, which is a 
very narrow resonance ($\Gamma <
10$ MeV) decaying into $D_s^+\pi^0$, is thought to be the missing
bound 
state with $J^P=0^+$
of the heavy-light system. This picture of $D_s(2317)$ composed of a 
heavy quark
and a light valence quark fits well with the heavy-quark, chiral
symmetries
that predict parity doubling states $(0^-,1^-)$ and $(0^+,1^+)$, with
the interparity mass splittings in the chiral limit given 
by the Goldberger-Treiman
relation \cite{zahed,bh,beh}. The subsequent observation of $1^+$ 
state 
$D_s(2460)$
\cite{cleo,barbar,belle} strongly 
supports this picture.

On the other hand, the two-quark picture of the resonances does not
play well
with the potential model calculations, which generally  predict
substantially
larger mass and width. According to the potential model calculation 
in Ref.
\cite{pe} the mass and width of $D_s(0^+)$ are, respectively, 2487 MeV
and a few 100 MeVs, with the width depending on
the light-quark axial coupling.
While the narrow decay width can be understood by the observed 
mass being below the
threshold of the strong decay channel $DK$ 
%, thus the channel kinematically blocked, 
and the isospin symmetry breaking, the substantially 
small observed mass is  puzzling.

Furthermore, this anomaly in the observed mass became more peculiar
when the Belle collaboration observed \cite{belle2}
non-strange $0^+$ state $D(2308)$,
whose mass is surprisingly close to $D_s(2317)$. The potential model
 predicts the mass splitting between these states to be $110$ MeV.
These peculiarities in the observed masses led to many models 
for the new
resonances, including, for example,  four-quark model 
\cite{cheng,terasaki},
 DK molecule models \cite{lipkin}, and unitarized meson 
 model \cite{rupp}. It is thus very important to clarify the nature
 of the newly discovered resonances.

The quoted numbers of the potential model calculation are based on 
Coulombic vector potential and a linear scalar potential. 
Modifications of the employed potentials
might remove the anomaly, but Cahn and Jackson \cite{cj} showed that, 
as far as the vector potential is kept Coulombic, it is unlikely that
the observed decay width and mass pattern of the resonances 
can be obtained from a potential model.

This suggests that the potential model be missing an essential physics
of the heavy-light system. Indeed, the conventional potential model
does not sufficiently take into account the chiral symmetry breaking nature of 
the QCD vacuum, with the chiral symmetry breaking encoded only in the 
light-quark constituent masses of the model. Since the light valence quark is 
chirally active
the heavy-light mesons can couple to the quantum fluctuations of the
Goldstone bosons of the 
QCD vacuum. This suggests that potential models must be
augmented by chiral radiative corrections. 
%Considering the success of 
%the quark model in hadrons as well as the success of the  potential model 
%in heavy quarkonia it may not be  unreasonable to expect the potential model, 
%when augmented by chiral radiative corrections, and possibly a modification
%of the potentials, would work reasonably well with the heavy-light 
%system as well.

In this paper we calculate chiral radiative corrections for the
bound state energies of the potential model, paying particular attention to the
mass splittings of the parity doubling states. Our main result is that chiral
corrections are large, comparable at least to $1/M_c$ corrections in $D$
mesons, where $M_c$
denotes the charm mass, rendering their inclusion to the potential model 
mandatory. 
%non-strange system they tend to widen the interparity mass gaps, 
%whereas for strange system they act in opposite direction.
Furthermore, for the parity doubling states, they tend to narrow the 
interparity mass gaps, and 
this effect is stronger in strange system than in nonstrange system,
with  a robust prediction of the mass ${\rm Gap}\equiv 
[m(D(0^+))-m(D(0^-))]-[m(D_s(0^+))-m(D_s(0^-))] \approx 90$ MeV (at
the axial coupling $g_A=0.82$) that is consistent with experiment.

%This behavior is consistent with the observed mass pattern of the parity 
%doubling  states in $D$ and $D_s$ system, and 
%opens the possibility to understand the small mass difference 
%between  $D_s(2317)$ and $D(2308)$ within a potential model.

%Whether a potential model with the chiral radiative corrections
%can provide a comprehensive understanding
%of the new resonances requires a global fitting of the model parameters.
%While the problem is interesting, in this note we do not pursue it
%and leave it for future investigation.

The potential model of heavy-light system \cite{roberts}
is based on the chiral quark model
\cite{georgi},
with the Lagrangian reading 
\bear
{\cal L} = \Psi^\dagger (i\partial_0 -H) \Psi
\eear
with $\Psi=(u,d,s)$ denoting the light quark fields and the Hamiltonian 
given by
\bear
H=H_0+\frac{1}{M} H_1 + \cdots %\frac{1}{M^2} H_2 +\cdots
\eear
where $M$ denotes the heavy quark mass.
The leading Hamiltonian $H_0$ in the heavy quark mass expansion reads
\bear
H_0= \gamma^0(-i \not\!\nabla + {\bf m}) + V(r)
\eear 
with the potential given in the form
\bear
V(r)=M + \gamma^0 V_s(r) +V_v(r)\,,
\eear
where $V_s$ and $V_v$ denote the scalar and vector potentials, respectively,
and  ${\bf m}=m_i\delta_{ij}$ denotes the constituent quark masses.
The energy spectra of resonances are obtained by solving the Dirac 
equation of $H_0$, followed by time-independent perturbations of the 
subleading terms. The free parameters of the model are fixed by
a global fitting of the predicted masses to those of the observed 
resonances. 

In this framework the chiral symmetry breaking of QCD is 
encoded only in the
constituent masses of the light quarks, and we shall see that
this is not sufficient enough. This inadequacy of the model can be easily 
remedied by noting that
the effective Hamiltonian is based on a truncated chiral quark model. 
In chiral quark model the light-quark--Goldstone boson interactions are
described by an infinite tower of derivative expansions, but
the term responsible for the one-loop corrections 
is the following axial coupling
%that induces the following correction term to $H_0$:
\bear
H_{\bar\psi\psi\pi}&=& -g_A \bar\Psi \not\!\! A \gamma_5 \Psi \nonumber\\
              &=& \frac{
g_A}{2f_{\pi}}\bar\Psi_i\gamma^\mu\gamma_5\Psi_j
	      \partial_\mu\Pi_{ij} +O(\Pi^2)
\label{chiral}
\eear
where $g_A$ is an axial coupling constant, and
\bear
A_\mu=\frac{i}{2}(\xi^\dagger\partial_\mu\xi-\xi\partial_\mu\xi^\dagger)
\eear
with $\xi=e^{i\Pi/2f_\pi}$, where $\Pi=\sum_{a=1}^{8}\pi^a \lambda^a$,  
$\lambda^a$ the Gell-Mann matrices, and $f_\pi=93$ MeV.
%, reads
%\be
%\Pi=\left(\begin{array}{ccc}
%\pi^0+\frac{\eta}{\sqrt{3}}&\sqrt{2}\pi^+&\sqrt{2}K^+ \\
%\sqrt{2}\pi^-&-\pi^0+\frac{\eta}{\sqrt{3}}&\sqrt{2}K^0\\
%\sqrt{2}K^-&\sqrt{2}\bar{K}^0&-\frac{2\eta}{\sqrt{3}}
%\end{array}\right) \,.
%\ee
%In this notation $f_\pi=93$ MeV.

%At this point 
We note that the inclusion of the axial 
term (\ref{chiral}) in the potential model Hamiltonian should not be 
unexpected, since this term was already employed in the
calculation of the 
decay widths in potential model. In general the widths,
which are the imaginary parts of the self-energies,
can be a few hundred MeVs, which indicates that the chiral radiative 
corrections to the resonance masses 
cannot be small, and so should be included in computation 
of the masses.

\begin{figure}
\resizebox{0.45\textwidth}{!}{
 \includegraphics{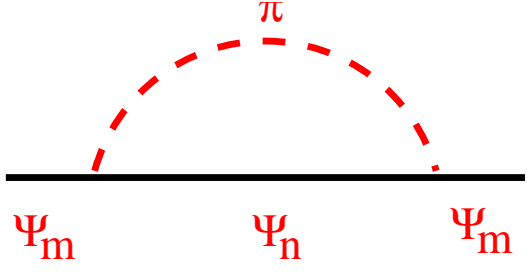}
}
\caption{One loop correction to the energy of the eigenstate $\Psi_{\bf m}$.}
\label{fig1}
\end{figure}

We shall now consider the corrections due to the chiral term (\ref{chiral})
to the energy of an eigenstate of $H_0$. Let us
denote the eigenenergy and normalized wavefunction  by
$E_{\bf m}$ and $\Psi_{\bf m}$, respectively. Here  ${\bf m}=\{n,l,j,m_j,q\}$ 
denotes the set of quantum numbers classifying the eigenstate of 
the light quark, with $n$, $q$, and  $l,j,m_j$ denoting the radial excitation,
quark flavor, and the angular momentum quantum numbers, respectively.
%under the potential provided by the heavy quark.
The correction to the energy $E_{\bf m}$ at one loop comes through the
diagram in Fig. \ref{fig1} and is given by
\bear
\Delta E_{\bf m}
%&=& \frac{ig_A^2}{4 f_\pi^2} \sum_{\bf n}\sum_\pi
%\zeta_{\pi}\int\frac{d^4k}{(2\pi)^4}
%\frac{j^\mu_{\bf mn}(\vec{k}){j^\nu_{\bf mn}}^\dagger(\vec{k})
%k_{\mu}k_\nu}{(E_{\bf m}-k_0-E_{\bf n}+i\epsilon) 
%(k^2-m_\pi^2+i\epsilon)}\nonumber\\
%&=&
=&&\frac{ig_A^2}{4 f_\pi^2} \sum_{\bf n}\sum_\pi\zeta_\pi \int\frac{d^4k}{(2\pi)^4}
\frac{|j_{\bf mn}(k)|^2}{(E_{\bf m}-k_0-E_{\bf n}+i\epsilon)
(k^2-m_\pi^2+i\epsilon)}
\label{loop}\eear
where 
\be
j_{\bf mn}(k)=
%j^\mu_{\bf mn}(\vec{k})k_\mu\equiv
\left(\int d^3\vec{x}
\,\,\Psi_{\bf m}^\dagger(\vec{x})\gamma^0
\gamma^\mu\gamma_5\Psi_{\bf n}(\vec{x})
e^{i\vec{k}\cdot\vec{x}} \right)k_\mu \,,
\ee
and $m_\pi$ denotes
the mass of the light meson exchanged and $\zeta_{\pi}$ 
represents the $SU(3)_{{\rm
flavor}}$ factors coming from the axial vertices.
%The wavefunction $\Psi_{\bf m}$ is normalized to
%\be
%\int d^3x\,\, \Psi_{\bf m}^\dagger(\vec{x})\Psi_{\bf m}(\vec{x})=1\,.
%\ee
%Here, for simplicity, the 
%$SU(3)_{\rm flavor}
%$ group factors at the axial vertices are suppressed for the moment, and
%will be taken into account in the final step.

Using the Dirac equations for the wavefunctions   
the current can be written as
\bear
%j_{\bf mn}(k)&=& k_0 \int d^3\vec{x}\,\,
%\Psi_{\bf m}^\dagger(\vec{x})\gamma_5\Psi_{\bf n}(\vec{x})
%e^{i\vec{k}\cdot\vec{x}} -% \nonumber \\ &&
%\int d^3\vec{x}\,\,\Psi_{\bf m}^\dagger(\vec{x})\gamma_5[
%E_{\bf m}-E_{\bf n}+\gamma^0(m_{\bf m}+m_{\bf n}+2 V_s)]
%\Psi_{\bf n}(\vec{x})e^{i\vec{k}\cdot\vec{x}}
% \nonumber \\
j_{\bf mn}(k)= (k_0-E_{\bf m}+E_{\bf n}) \rho_{\bf mn}^{(1)}(\vec{k})+ 
\rho_{\bf mn}^{(2)}(\vec{k})\,,
\label{current}\eear
where
\bear
\rho_{\bf mn}^{(1)}(\vec{k})&=&\int \,\,d^3\vec{x}\,\,
\Psi_{\bf m}^\dagger(\vec{x})
\gamma_5\Psi_{\bf n}(\vec{x})
e^{i\vec{k}\cdot\vec{x}}\,, \nonumber\\%\quad
\rho_{\bf mn}^{(2)}(\vec{k})&=&\int \,\,d^3\vec{x}\,\,
\Psi_{\bf m}^\dagger(\vec{x})\gamma^0
\gamma_5(m_{\bf m}+ m_{\bf n}+2 V_s)\Psi_{\bf n}(\vec{x})e^{i\vec{k}\cdot\vec{x}}\,.
\label{rhodef}
\eear
%Substituting (\ref{current}) into (\ref{loop}) we obtain 
%\bear
%\Delta E_{\bf m}&=& \frac{-i g_A^2}{4 f_\pi^2} \sum_{\bf n}\sum_\pi
%\zeta_{\pi}\int\frac{d^4k}{(2\pi)^4}
%\left[
%\frac{(k_0+E_{\bf n}-E_{\bf m})|
%\rho_{\bf mn}^{(1)}(\vec{k})|^2}{k^2-m_\pi^2+i\epsilon}
%+\frac{2 {\rm Re}[\rho_{\bf mn}^{(1)}(\vec{k})
%\rho_{\bf mn}^{(2)}(\vec{k})^*]}{
%k^2-m_\pi^2+i\epsilon} + \right.\nonumber \\ 
%&&\left.\frac{|\rho_{\bf mn}^{(2)}(\vec{k})|^2}{
%(k_0-E_{\bf m}+E_{\bf n}-i\epsilon)(k^2-m_\pi^2+i\epsilon)}\right]\,.
%\label{dele00}\eear
Substituting (\ref{current}) into (\ref{loop}), and performing the 
integration over $k_0$ we obtain
\bear
\Delta E_{\bf m}=\sum_{\bf n}\sum_\pi\zeta_{\pi} J({\bf m},{\bf n},m_\pi)
\label{delE0}
\eear
where
\bear
J({\bf m},{\bf n},m_\pi)&=&
\frac{ -g_A^2}{8 f_\pi^2} \int\frac{d^3k}{(2\pi)^3 E_\pi(\vec{k})}
\left[(E_{\bf n}-E_{\bf m})|\rho_{\bf mn}^{(1)}(\vec{k})|^2
+\right. \nonumber \\
&&\hspace{-1.5cm} \left. 2 {\rm Re}[\rho_{\bf mn}^{(1)}(\vec{k})\rho_{\bf
mn}^{(2)}(\vec{k})^*] + \frac{|\rho_{\bf mn}^{(2)}(\vec{k})|^2}{
E_\pi(\vec{k})-E_{\bf m}+E_{\bf n}-i\epsilon}
\right] 
\label{delE}
\eear
with $E_\pi(\vec{k})=\sqrt{\vec{k}^2+m_\pi^2}$.
%Note that, formally, the integration over $k_0$ in the first term in
%(\ref{dele00}) is divergent, but this problem can be easily removed by
%taking a symmetric cutoff and then letting the cutoff go infinity.
%Taking this symmetric cutoff is necessary since, otherwise, an unphysical
%imaginary part is induced.

%When $E_{\bf m} > E_{\bf n}+m_\pi$ the last term can be 
%singular and gives rise to an imaginary part to $\Delta E_{\bf m}$, which
%is nothing but the decay width for 
%$\Psi_{\bf m}\to \Psi_{\bf n}+\pi$. For the correction 
%to the bound state energy the real part is to be taken.
%The partial decay width
%$\Gamma=-2{\rm Im}\, \zeta_\pi J({\bf m},{\bf n})$ for $\Psi_{\bf m}\to
%\Psi_{\bf n}+\pi$ is given by
%\be
%\Gamma=\frac{g_A^2 \zeta_\pi
%|\rho_{\bf mn}^{(2)}(p_0)|^2 p_0}{8\pi f_\pi^2}, \quad 
%p_0=\sqrt{(E_{\bf m}-E_{\bf n})^2-m_\pi^2} \,,
%\ee
%which agrees, in the heavy quark limit, with Ref. \cite{pe}.

We shall now write the currents $\rho_{\bf mn}^{(i)}$ in terms of the radial
functions of the eigenfunctions, which can be written as 
\begin{equation}
   \Psi_{\bf m}(\vec{r}) = 
   \left( 
     \begin{array}{c}
        i f_{n\ell j q}(r) \\
        g_{n\ell j q}(r) \vec{\sigma}\cdot\hat{r}
     \end{array}
   \right) 
   {\cal Y}_{\ell j m}(\hat{r}),
\end{equation}
where ${\cal Y}_{\ell j m}(\hat{r})$ is the spinor harmonics. 
%the eigenfunctions of the total angular momentum operator of the light 
%quark $\vec{J}=\vec{L} + \frac12\vec{\sigma}$. Explicitly, they are
%\begin{eqnarray}
%  {\cal Y}_{\ell j m}
%   &=& 
%   \sqrt{\frac{\ell+m+\frac12}{2\ell+1}} 
%      Y_{\ell, m-\frac12}(\hat{r}) \chi_{+}
%   +
%   \sqrt{\frac{\ell-m+\frac12}{2\ell+1}} 
%      Y_{\ell, m+\frac12}(\hat{r}) \chi_{-}, 
%   \hskip 2em (j=\ell+\textstyle\frac12),
%  \label{Y+}
%\\
%  {\cal Y}_{\ell j m}
%   &=& 
%   \sqrt{\frac{\ell-m+\frac12}{2\ell+1}} 
%      Y_{\ell, m-\frac12}(\hat{r}) \chi_{+}
%   -
%   \sqrt{\frac{\ell+m+\frac12}{2\ell+1}} 
%      Y_{\ell, m+\frac12}(\hat{r}) \chi_{-}, 
%   \hskip 2em (j=\ell-\textstyle\frac12).
%  \label{Y-}
%\end{eqnarray}
%where 
%$$
%   \chi_{+} = \left( \begin{array}{c} 1 \\ 0 \end{array}\right), 
%   \hskip 2em
%   \chi_{-} = \left( \begin{array}{c} 0 \\ 1 \end{array}\right). 
%$$
%The phases of the Clebsh-Gordan coefficients used in eqs.(\ref{Y+})
%and(\ref{Y-}) are chosen to satisfy a convenient relation 
%\begin{equation}
%(\vec{\sigma}\cdot\hat{r}) {\cal Y}_{\ell=j\pm\frac12,jm}(\hat{r}) 
%= {\cal Y}^{(\mp)}_{\ell^\prime=j\mp \frac12,jm}(\hat{r}).
%\end{equation}
Since 
the light quark wavefunctions are eigenstates of the angular 
momentum operator, 
it is convenient to expand the plane wave $\exp(i\vec{k}\cdot\vec{r})$
in the definition of the currents $\rho^{(1,2)}_{\bf mn}$ in
(\ref{rhodef}) as 
\begin{equation}
  e^{i\vec{k}\cdot\vec{r}} 
  = 4\pi \sum_{\ell=0}^\infty i^\ell j_\ell^{}(kr) 
  \sum_{m=-\ell}^{+\ell} Y^*_{\ell,m}(\hat{r}) Y_{\ell,m}(\hat{k})\,. 
  \label{expansion}
\end{equation}
Then, the currents $\rho^{(1,2)}_{\bf mn}$ can be expanded as  
\begin{equation}
    \rho^{(1,2)}_{\bf mn}(\vec{k}) = 
      \sum_{\ell,m} \rho^{(1,2)}_{\bf mn}(|\vec{k}|,\ell,m) 
      Y_{\ell,m}(\hat{k}) 
      \label{e22}
\end{equation}
with, up to a common phase, 
%\begin{eqnarray}
%   \rho^{(1)}_{\bf mn}(|\vec{k}|,\ell,m) 
%   &=& 
%     -i^{\ell+1}_{} 4\pi \int^\infty_0 r^2 dr 
%       (f^{}_{\bf m}(r) g^{}_{\bf n}(r) 
%        -f^{}_{\bf n}(r) g^{}_{\bf m}(r)) j_\ell^{}(kr)  
% \nonumber\\
%   && \hskip 5em
%       \times \int d\Omega_r  
%        {\cal Y}^{\dagger}_{\bf m} Y^*_{\ell,m}(\hat{r})
%        \vec{\sigma}\cdot\hat{r} {\cal Y}^{}_{\bf n}(\hat{r}),
%\\
%   \rho^{(2)}_{\bf mn}(|\vec{k}|,\ell,m) 
%   &=& 
%     -i^{\ell+1}_{} 4\pi \int^\infty_0 r^2 dr 
%       (f^{}_{\bf m}(r) g^{}_{\bf n}(r) 
%        +f^{}_{\bf n}(r) g^{}_{\bf m}(r)) j_\ell^{}(kr)
%        (m_{\bf m} + m_{\bf n} + 2V_s)
% \nonumber\\
%   && \hskip 5em
%       \times \int d\Omega_r  
%        {\cal Y}^{\dagger}_{\bf m} Y^*_{\ell,m}(\hat{r})
%        \vec{\sigma}\cdot\hat{r} {\cal Y}^{}_{\bf n}(\hat{r}).
%\end{eqnarray}
%
%Using the Wigner-Eckart theorem on the spherical harmonics the $m$-dependence
%of the matrix elements can be factored out as
\begin{eqnarray}
   \rho^{(1,2)}_{\bf mn}(|\vec{k}|,l_\pi,m_\pi) 
   &=& 
      4\pi \int^\infty_0 r^2 dr 
       \tilde \rho_{\bf mn}^{(1,2)}(r) j_{l_\pi}^{}(kr)  \langle j m_j l_\pi m_\pi | j^\prime m_j^\prime\rangle 
         \langle \ell j ||Y^*_{l_\pi} \vec{\sigma}\cdot\hat{r} 
         || \ell^\prime j^\prime\rangle \nonumber \\
	 &\equiv& 
       \langle j m_j l_\pi m_{\pi} | j^\prime m_j^\prime\rangle 
       \rho^{(1,2)}_{\bf mn}(|\vec{k}|,l_\pi) \,,
\label{selection}
\eear
where
\bear
\hspace{-0.5cm}\tilde \rho_{\bf mn}^{(1)}&=&f^{}_{\bf m}(r) g^{}_{\bf n}(r) 
        -f^{}_{\bf n}(r) g^{}_{\bf m}(r)\,, \nonumber\\% \quad
\hspace{-0.5cm}	\rho_{\bf mn}^{(2)}&=& \big(f^{}_{\bf m}(r) g^{}_{\bf n}(r) 
       \! +\!f^{}_{\bf n}(r) g^{}_{\bf m}(r)\big)(m_{\bf m} \!+\! m_{\bf n}
	\!+\! 2V_s)\,.
\label{YYY}
\eear
%\begin{eqnarray}
%   \rho^{(1,2)}_{\bf mn}(|\vec{k}|,\ell,m) 
%   &=& 
%     -i^{\ell+1}_{} 4\pi \int^\infty_0 r^2 dr 
%       (f^{}_{\bf m}(r) g^{}_{\bf n}(r) 
%        -f^{}_{\bf n}(r) g^{}_{\bf m}(r)) j_\ell^{}(kr)  
% \nonumber\\
%   && \hskip 5em
%       \times \langle j m_j \ell m | j^\prime m_j^\prime\rangle 
%         \langle \ell j ||Y^*_\ell \vec{\sigma}\cdot\hat{r} 
%         || \ell^\prime j^\prime\rangle
%\nonumber\\
%  &\equiv &
%       \langle j m_j \ell m | j^\prime m_j^\prime\rangle 
%       \rho^{(2)}_{\bf mn}(|\vec{k}|,\ell)\,, \nonumber \\
%   \rho^{(2)}_{\bf mn}(|\vec{k}|,\ell,m) 
%   &=& 
%     -i^{\ell+1}_{} 4\pi \int^\infty_0 r^2 dr 
%       (f^{}_{\bf m}(r) g^{}_{\bf n}(r) 
%        +f^{}_{\bf n}(r) g^{}_{\bf m}(r))(m_{\bf m} + m_{\bf n} + 2V_s)
%	j_\ell^{}(kr)  
% \nonumber\\
%   && \hskip 5em
%       \times \langle j m_j \ell m | j^\prime m_j^\prime\rangle 
%         \langle \ell j ||Y^*_\ell \vec{\sigma}\cdot\hat{r} 
%         || \ell^\prime j^\prime\rangle
%\nonumber\\
%  &\equiv &
%       \langle j m_j \ell m | j^\prime m_j^\prime\rangle 
%       \rho^{(2)}_{\bf mn}(|\vec{k}|,\ell)\,.
%       \label{YYY}
%\end{eqnarray}
%and a similar equation for $\rho^{(2)}_{\bf mn}(|\vec{k}|,\ell,m)$.
Here $\{ \ell, j, m_j \}$ are the angular momentum quantum numbers of the 
state ${\bf m}$, $\{ \ell^\prime, j^\prime, m^\prime_j \}$ 
 those of the state ${\bf n}$ and $\ell_\pi$ is the angular momentum 
of the intermediate meson appearing in the expansion 
(\ref{expansion}), and
$\langle \ell j 
||Y^*_{l_\pi} \vec{\sigma}\cdot\hat{r} 
||\ell^\prime j^\prime\rangle$
denotes the reduced matrix element.
%which can be obtained by  evaluating 
%the angular integral and then  factoring out the corresponding 
%Clebsh-Gordan coefficient as 
%\begin{equation}
%   \int d\Omega_r  
%      {\cal Y}^{\dagger}_{\ell j m_j} Y^*_{\ell_\pi,m}(\hat{r})
%        \vec{\sigma}\cdot\hat{r} 
%       {\cal Y}^{}_{\ell^\prime,j^\prime,m_j^\prime}(\hat{r}),
%   \equiv 
%    \langle j m_j \ell_\pi m | j^\prime m_j^\prime\rangle 
%    \langle \ell j ||Y^*_{\ell_\pi} \vec{\sigma}\cdot\hat{r} 
%         ||\ell^\prime j^\prime\rangle.
%\end{equation}
Eq.(\ref{selection}) provides a selection rule for the possible intermediate 
light meson angular momentum, $l_\pi$, for a given internal state and vice
versa. 

Now, % performing first the $k^0$-integration in (\ref{loop}) and
doing the angular part of the $\vec{k}$- integration, 
% in the energy correction %(\ref{delE}) 
which can be easily 
carried out with the decomposition (\ref{e22}), and  using the unitarity 
relation 
\begin{equation}
   \sum_{m_j^\prime}
      \sum_{m_\pi+m_j=m_j^\prime} 
       % \big( 
           \langle j m_j l_\pi m_\pi | j^\prime m_j^\prime \rangle^2
        %\big)^2 
      = \sum_{m_j^\prime} \left\{ 1 \right\} = 2j^\prime+1, 
\end{equation}
we can rewrite the loop corrections to the energy as 
\begin{equation}
   \Delta E_{\bf m} = \sum_{\bf n}\sum_{\pi,l_\pi} \zeta_\pi 
                      J({\bf m},{\bf n},l_\pi) 
		      \frac{2j_{\bf n}+1}{2j_{\bf m}+1},  
  \label{E_cor_new}
\end{equation}
where
\begin{eqnarray}
%  && J({\bf m},{\bf n},\pi(\ell))
%\nonumber\\
%  && \hskip 1em
 J({\bf m},{\bf n},l_\pi)
    &=& -\frac{g_A^2}{8f_\pi^2}  
      \int \frac{k^2 dk}{(2\pi)^3 E_\pi} 
   \left[ 
      (E_{\bf n} - E_{\bf m})|\rho^{(1)}_{\bf mn}(|\vec{k}|,l_\pi)|^2
   \right.\nonumber\\
 &&\left.%\hskip 5em
     \hspace{-2.3cm}+ 2{\rm Re}[\rho^{(1)}_{\bf mn}(|\vec{k}|,l_\pi)
            \rho^{(2)*}_{\bf mn}(|\vec{k}|,l_\pi)]
     + \frac{|\rho^{(2)}_{\bf mn}(|\vec{k}|,l_\pi)|^2}
            {E_\pi - E_{\bf m} + E_{\bf n}-i\epsilon}
   \right].
  \label{J_mn_lm}
\end{eqnarray}

We now focus on the energy corrections for the lowest energy
parity doubling states, \[D(0^-),D_s(0^-) 
\quad {\rm and}\quad D(0^+),D_s(0^+)\,.\]
% $0^-_{d,s}, 0^+_{d,s}$, 
They have the quantum numbers 
${\bf m}=\{1,0,\frac{1}{2},\pm\frac{1}{2},q=(d,s)\}$,
$\{1,1,\frac{1}{2},\pm \frac{1}{2},q=(d,s)\}$,
and, to shorten the notations, will be denoted by
${\bf 0}_{d,s},{\bf 1}_{d,s}$, respectively. For these states 
$\rho^{(i)}_{{\bf m},{\bf n}}(|\vec{k}|,\ell_\pi)$ in (\ref{J_mn_lm}) are
given by
\begin{eqnarray}
   \rho^{(1)}_{{\bf m},{\bf n}}(|\vec{k}|,\ell_\pi) 
  &=&      \sqrt{4\pi} \int^\infty_0 r^2 dr 
       (f^{}_{{\bf m}}(r) g^{}_{{\bf n}}(r) - f^{}_{{\bf n}}(r) g^{}_{{\bf m}}(r)) j_{\ell_\pi}^{}(kr),\\
   \rho^{(2)}_{{\bf m},{\bf m}}(|\vec{k}|,\ell_\pi) 
   &=& 
      \sqrt{4\pi} \int^\infty_0 r^2 dr 
       (f^{}_{{\bf m}}(r) g^{}_{{\bf n}}(r) 
       +	 f^{}_{{\bf n}}(r) g^{}_{{\bf m}}(r)) 
% \hskip 7em \times 
(m_{\bf m} + m_{\bf n} + 2V_s)j_{\ell_\pi}^{}(kr)\,.
\end{eqnarray}

Before giving the numerical result, we comment on the divergence of the 
loop corrections.
The loop correction $J({\bf m}, {\bf n},l_\pi)$
for given ${\bf m,n}$ is free from ultraviolet (UV)
divergence, with the wavefunctions providing the UV cutoff. 
However, the total loop correction obtained by summing over the internal
states is quadratically divergent. The quadratic divergence comes from
the first two terms of (\ref{delE}), which can be
easily summed over the internal states using the Dirac equation for
the wavefunctions and 
%seen  by
%performing the summation over the internal states 
%first in (\ref{delE}) 
%before doing the momentum integration.
%Then,
 the definition of $\rho_{\bf mn}^{(i)}$ in (\ref{rhodef}).
% the second term in (\ref{delE}) can be summed easily, and
%using the Dirac equation the first term also can be summed, which
This gives the sum of the first two terms as 
\bear
\Delta E_{\bf m}^{\rm quad. div.}&=&\sum_\pi \zeta_\pi
\frac{ -g_A^2}{8 f_\pi^2}\left[\int\frac{d^3k}{(2\pi)^3
E_\pi(\vec{k})}\right] \int d^3 x\,\,\Psi_{\bf m}
^\dagger \gamma^0(m_{\bf m}+m_{\bf n}+ 2 V_s)\Psi_{\bf m}\,.
\eear
The third term in (\ref{delE}) is at most linearly divergent.
%momentum independent when summed over the internal states at a fixed
%light-quark flavor, which immediately shows that the
%integration over the momentum is quadratically divergent. 
%It is easy to check that any UV divergences in the
%remaining two terms have the same sign as that in the second term, and
%so the quadratic divergence cannot be canceled.
%Therefore, the chiral loop correction to the energy is quadratically
%divergent.
This quadratic divergence of the energy correction is not unexpected
since the chiral quark model is an effective theory
valid only at low energies.
%, and the loop corrections should become unreliable
%when they are between a low lying state
%and a highly excited state.
To regularize the UV divergence we introduce 
%one may introduce a cutoff in the momentum, or truncate the summation over 
%the internal states at a certain energy level.
%Since chiral loop corrections
%should be unreliable at high momentum the former approach is obviously
%more desirable, and so 
%In this paper we take the first approach and  introduce 
a three-momentum cutoff of the form
\bear
e^{-\vec{k}^2/\Lambda_{\rm UV}^2}
\eear
to the integrand in (\ref{J_mn_lm}). 
%where $\Lambda_{\text{UV}}$ is a constant.
We regard $\Lambda_{\rm UV}$ as the physical cutoff of
the chiral quark model, but we shall see that our 
main result on the mass Gap
depends little on the cutoff.

%Now the correction to the energy is well defined but it depends on 
%the free parameter $\Lambda_{\text{UV}}$.

%Because of the severe UV divergence and the arbitrary cutoff
%the chiral loop corrections may not be considered as a quantitatively 
%accurate 
%prediction of the chiral effects in heavy-light mesons, but, nevertheless,
%one may expect that they can reveal the qualitative nature  of 
%the chiral effects, and  shed lights
%on some of the peculiar properties of the recently observed resonances.
%In the following we apply the one loop corrections to the lowest
%energy parity doubling states, $D(0^-),D(0^+)$ and $D_s(0^-),D_s(0^+)$.

To obtain the eigenfunctions of $H_0$ we should first fix the parameters 
of the model. In the following we shall follow the setup as well as
use the 
parameter values given in Ref. \cite{pe}, in which the vector 
potential is Coulombic and the scalar potential is a linearly confining 
potential. 
The model has nine free parameters that are to be fixed by a global 
fitting of the predicted resonance masses
%, which also include $1/M$ corrections,
to those observed values. For details we refer the readers
to the above reference. Of course, the parameters were fixed 
without taking the chiral corrections into account, but we can use those
values  to estimate the 
loop correction effects at leading order, which are comparable in magnitude
to the  $1/M_c$ corrections
and so play only a subleading role in fitting the parameters.
We also need to fix the axial coupling $g_A$,
which is a free parameter in the chiral quark model. In principle it can
be determined by fitting the hadronic decay width of an excited heavy-light
meson to experimental data, or by lattice simulation. These approaches
estimate $g_A$ to be around unity with a considerable uncertainty \cite{pe}.

%In this paper we put $g_A=1$, for simplicity, and the corrections at other
%values of $g_A$ can be obtained by a trivial scaling.

It is convenient to organize the energy corrections in terms of 
the angular momentum
$l_\pi$ of the intermediate light mesons. For a given $l_\pi$ we sum 
over all allowed internal states, which can  be selected by the angular momentum
and parity conservations at the axial vertex, up to the first 10 radial
excitations. As can be seen in Table \ref{ndep} the corrections
drop rapidly at higher radial excitations. 
Our result is summarized in Table \ref{ldep} at varying 
cutoffs $\Lambda_{\rm UV}$. 
At a smaller cutoff
$\Lambda_{\rm UV}=700$ MeV the 
corrections drop quickly as $l_\pi$ increases,
whereas at a larger cutoff $\Lambda_{\rm UV}=1200$ MeV they 
drop slowly, reflecting the
UV divergence of the chiral corrections. At all the 
cutoffs considered, the largest
contributions come from the 
$l_\pi=1$ modes, and at larger cutoffs contributions
from $l_\pi$ as large as four are significant. 

Not surprisingly, the total energy corrections depend strongly on the
UV cutoff. At $\Lambda_{\rm UV}=700$ MeV they are 
a few hundred MeVs but at
$\Lambda_{\rm UV}=1200$ they are in GeV order.
This shows that in our model the physical cutoff should be about 700 MeV.
%This clearly shows that
%the absolute values for the corrections we obtain cannot be considered as
%a quantitative prediction of the chiral effects in heavy-light mesons. 
Although the total corrections are sensitive on the cutoff,
%However, what we are more interested in is the interparity
%mass gap between the
%parity doubling, strange and nonstrange  states. We would expect the mass 
%gap is less dependent on the cutoff. 
we expect the difference of the interparity mass gaps between the
parity doubling states is less sensitive on the cutoff. 
Indeed, summing the contributions
up to $l_\pi=9$ we find that
\bear
\Delta E{\bf 1}_d-\Delta E{\bf 0}_d &=&-146, -312, -447 \quad {\rm MeV}\,, 
\nonumber \\ 
\Delta E{\bf 1}_s-\Delta E{\bf 0}_s &=&-271, -442, -579  \quad {\rm MeV}\,,
\label{gap0}
\eear
and
\bear
{\rm Gap}&\equiv& (\Delta E{\bf 1}_d-\Delta E{\bf 0}_d)-
(\Delta E{\bf 1}_s-\Delta E{\bf 0}_s)\nonumber \\
&=&125, 130, 132 \quad {\rm MeV}
\label{gap}
\eear 
at $\Lambda_{\rm UV}=700,1000,1200$ MeV,
respectively, and $g_A=1$.
%Except for the case at
%$\Lambda_{\rm UV}=700$ these
%numbers are expected to get modified when higher $l_\pi$ modes are taken
%into account.  
This  shows that
the chiral corrections shrink the interparity gaps, both in strange
and nonstrange systems, but do so more in the 
strange system. This may be an
explanation for the unusually smaller mass of $D_s(2317)$ 
than given in the potential model. 
More interestingly, the mass Gap is remarkably stable under
variation of the cutoff.
This suggests that the Gap in Eq. (\ref{gap}) comes almost entirely 
from the low energy region far down the cutoff.
To see this we plot in Fig. 2 the differential Gap ${\cal G}(k)$,
defined as the Gap before
the integration over the momentum variable $k$, that is,
\[{\rm Gap}\equiv \int_0^\infty {\cal G}(k) d k\,,\]
which can be
obtained from the proper combination of the integrands for the
energy corrections given
in Eq. (\ref{J_mn_lm}). The discontinuity in the plot comes from
our implementation of the principal value prescription for the last term
in the integrand in Eq. (\ref{J_mn_lm}) which has a pole at $
k_p=\sqrt{(E_{\bf m}-E_{\bf n})^2 -m_\pi^2}$ when
the external state is ${\bf 1}_d$ and internal state is 
${\bf 0}_d$ ($E_{\bf m}=2282 \,{\rm MeV}$, $E_{\bf n}=1895\, {\rm MeV}, 
m_\pi=140 \,{\rm MeV}$). Our numerical 
code handles the principal value integration
using the identity 
\be
\int_0^\infty \frac{f(k)}{k-k_p} d k = \int_0^{2 k_p}
\frac{f(k)-f(k_p)}{k-k_p} d k +\int_{2 k_p}^\infty 
\frac{f(k)}{k-k_p} d k
\ee
which is valid for any smooth function $f(k)$. With this 
the pole at $k_p$ is now
removed from the integrand and there appears a discontinuity
at $2 k_p$. Notice that the differential Gap has a peak 
around $k\approx 250$ MeV and the bulk of the contribution to the Gap
comes from the low energy region. This shows that our prediction of 
the Gap is not affected by the UV physics and thus safe from
the truncation of the higher derivative terms 
in the chiral quark model.

\begin{figure}
\resizebox{0.48\textwidth}{!}{
 \includegraphics{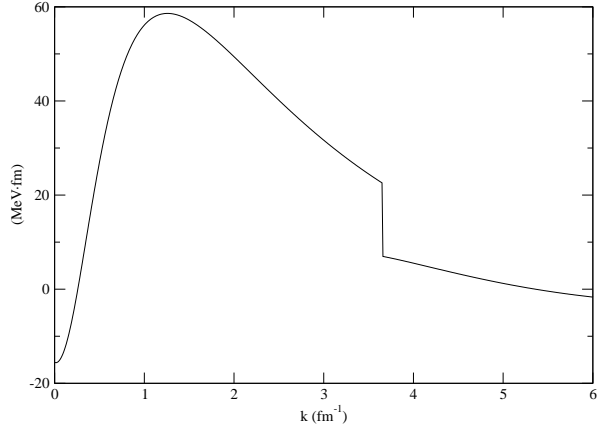}
}
\caption{The differential Gap ${\cal G}(k)$. 
No cutoff applied ($\Lambda_{\rm
UV}=\infty$) and $g_A=1$. Notice
that the bulk of the contribution to the
Gap comes from the low energy region.}
\label{fig2}
\end{figure}

%, and is consistent with experiment.
%Although the numbers in (\ref{gap0}) are expected to get modified 
%when higher $l_\pi$ modes are taken
%into account, the mass gap in  (\ref{gap}) is affected little by the 
%higher $l_\pi$ modes.  
%\bear
%(\Delta E{\bf 1}_d-\Delta E{\bf 0}_d)-
%(\Delta E{\bf 1}_s-\Delta E{\bf 0}_s)=125, 130, 132 \quad \text{MeV}
%\label{gap}
%\eear 
%at $\Lambda_{\rm UV}=700,1000,1200$ MeV, respectively. 
%Note that these numbers are 
%much less dependent on the cutoff, and affected little by the higher $l_\pi$
%modes ignored.  
%Moreover, the numbers are
%consistent with the observed data that shows larger interparity gap
%for nonstrange system. 
Without the chiral corrections the potential model
predicts almost vanishing Gap, %
%$(E{\bf 1}_d-E{\bf 0}_d)-(E{\bf 1}_s-E{\bf 0}_s)$, 
while experimentally
 it is about $95$ MeV.
If we take $g_A=0.82$, as given in Ref. \cite{pe},  we get experimentally
consistent 90 MeV for the Gap. We note that a similar result was 
observed by Eichten
using the heavy-light chiral Lagrangian \cite{eichten}, but
this was obtained without taking into account the tree-level
mass terms for the heavy-light mesons arising from the
explicit chiral
symmetry breaking. 

It is notable that the Gap is dominated by the contributions
from the $n=1, l_\pi=0$ modes, as can be seen from the 
Tables \ref{ndep},\ref{ldep}. Interestingly, these modes widen the
interparity mass gap in nonstrange system whereas in strange system narrow
the gap, but still the Gap from these modes are already consistent with
experiment. 

In conclusion, we calculated the 
one-loop chiral corrections for the heavy-light
mesons in potential model based on the truncated chiral quark model,
and have 
shown that the chiral corrections can account 
for the unusually small mass of $D_s(2317)$ and the narrow mass difference
between $D_s(2317)$ and $D(2308)$. 
Our calculation strongly supports the two-quark picture of
the new resonances composed of a heavy quark 
and a light valence quark.

%\begin{center}
%\begin{table*}
%\caption{\label{ndep}Energy corrections from the first five radial excitations 
%at $l_\pi=0$. $\text{Gap}\equiv (\Delta E{\bf 1}_d-\Delta E{\bf 0}_d)-
%(\Delta E{\bf 1}_s-\Delta E{\bf 0}_s)$. Values
%are at $\Lambda_{\rm UV}=1200$ MeV and $g_A=1$. Units are in MeV.}
%\begin{ruledtabular}
%\begin{tabular}{cccccc} 
%        $n$      &1&2&3&4&5 \\ \hline
%$ \Delta  E{\bf 0}_d$ &-124& -46& -18&  -7&  -3 \\ \hline
% $ \Delta E{\bf 1}_d$  & 109&-110& -49& -22& -10 \\ \hline
%                 
%  $ \Delta E{\bf 0}_s$  &-148& -54& -21&  -9&  -4 \\ \hline
%  $ \Delta E{\bf 1}_s$  & -43&-110& -52& -25& -12 \\ \hline
%  Gap                   &128&-8&0&1&1\\
%\end{tabular}
%\end{ruledtabular}
%\end{table*}
%\end{center}

\begin{center}
\begin{table*}
\caption{\label{ndep}Energy corrections from the first five radial excitations 
at $l_\pi=0$. $\text{Gap}\equiv (\Delta E{\bf 1}_d-\Delta E{\bf 0}_d)-
(\Delta E{\bf 1}_s-\Delta E{\bf 0}_s)$. Values
are at $\Lambda_{\text{UV}}=700$ MeV and $g_A=1$. Units are in MeV.}
\begin{ruledtabular}
\begin{tabular}{cccccc} 
        $n$      &1&2&3&4&5 \\ \hline
$ \Delta  E{\bf 0}_d$ &-23& -3& 0&  0&  0 \\ \hline
 $ \Delta E{\bf 1}_d$  & 29&-49& -8& -1& 0 \\ \hline
                 
  $ \Delta E{\bf 0}_s$  &-33& -3& 0&  0&  0 \\ \hline
  $ \Delta E{\bf 1}_s$  & -108&-49& -9& -1& 0 \\ \hline
  Gap                   &127&0&1&0&0\\
\end{tabular}
\end{ruledtabular}
\end{table*}
\end{center}

%\begin{center}
\begin{table*}
\caption{\label{ldep}Energy corrections ($g_A=1$) vs $l_\pi$. 
$\text{Gap}\equiv (\Delta E{\bf 1}_d-\Delta E{\bf 0}_d)-
(\Delta E{\bf 1}_s-\Delta E{\bf 0}_s)$.
Units are in MeV. }
\begin{ruledtabular}
\begin{tabular}{clrrrrrrrrrr} 
            & $l_{\pi}$& 0&1&2&3&4&5&6&7&8&9 \\ \hline 
            & $\Lambda_{\text{UV}}=700$& -26&-142&-50&-16&-5&-1&0&0&0&0 \\ 
$\Delta E{\bf 0}_d$ &
$\Lambda_{\text{UV}}=1000$&-116&-361&-196&-93&-41&-17&-7&-3&-1&0\\ 
	    &
$\Lambda_{\text{UV}}=1200$&-201&-561&-362&-202&-105&-52&-25&-12&-6&-3 \\ 
\hline
	    & $\Lambda_{\text{UV}}=700$&-29&-192&-101&-42&-15&-5&-2&0&0&0
	    \\ 
$\Delta E{\bf 1}_d$ &
$\Lambda_{\text{UV}}=1000$&-64&-400&-310&-185&-98&-49&-23&-11&-5&-2 \\ 
	    &
$\Lambda_{\text{UV}}=1200$&-93&-558&-502&-349&-215&-124&-69&-37&-19&-10 
\\ \hline
            & $\Lambda_{\text{UV}}=700$& -37&-136&-47&-14&-4&0&0&0&0&0 \\ 
$\Delta E{\bf 0}_s$ &
$\Lambda_{\text{UV}}=1000$&-141&-376&-197&-88&-37&-15&-6&-2&0&0 \\ 
	    &
$\Lambda_{\text{UV}}=1200$&-239&-599&-372&-197&-97&-46&-21&-10&-4&-2
\\ \hline
            & $\Lambda_{\text{UV}}=700$& -168&-186&-98&-39&-14&-4&-1&0&0&0 
	    \\ 
$\Delta E{\bf 1}_s$ &
$\Lambda_{\text{UV}}=1000$&-217&-413&-317&-183&-94&-45&-20&-9&-4&-2\\
            & $\Lambda_{\text{UV}}=1200$
	    &-253&-589&-524&-354&-210&-117&-63&-32&-16&-8\\ 	\hline
            &$\Lambda_{\text{UV}}=700$&128&0&0&-1&0&0&-1&0&0&0\\
  Gap       &$\Lambda_{\text{UV}}=1000$&128&-2&6&3&0&-2&-2&-1&0&0 \\
	    &$\Lambda_{\text{UV}}=1200$&122&-7&12&10&3&-1&-2&-3&-1&-1\\

\end{tabular}
\end{ruledtabular}
\end{table*}

\begin{acknowledgments}
T.L. is thankful to H.-Y. Cheng, P. Ko and M. Rho for
useful discussions, and especially to E. Eichten for valuable 
suggestions
and comments, and supported in part by 
Korea Research Foundation 
Grant (KRF-2005-015-C00107,KRF-2004-015-C00095) and
research fund from Kunsan National University.
B-Y.P. is grateful for the hospitality of the Special 
Research Centre for
the Subatomic Structure of Matter at the University of Adelaide,
where part of this work has been done.
\end{acknowledgments}

%\bibliographystyle{apsrev}
%\bibliographystyle{JHEP}
%\bibliography{dsdu1}

\providecommand{\href}[2]{#2}\begingroup\raggedright\endgroup

%\end{thebibliography}

\end{document}